\documentclass[12pt]{iopart}

\usepackage{graphicx}
\usepackage{bm}
\usepackage{iopams}

\begin{document}

\title{Elasticity of smectic liquid crystals with focal conic domains}

\author{S. Fujii$^1$, S. Komura$^2$, Y. Ishii$^3$\footnote{Present address: Department of Physics, Kyoto University, Kyoto 606-8502, Japan}, C.-Y. D. Lu$^4$
}

\address{
$^1$ Department of Chemistry, 
Nagaoka University of Technology,
Niigata 940-2188, Japan\\
$^2$ Department of Chemistry, 
Tokyo Metropolitan University, Tokyo 192-0397, Japan\\
$^3$ Department of Organic and Polymeric Materials, 
Tokyo Institute of Technology, Tokyo 152-8552, Japan\\
$^4$ Department of Chemistry, 
National Taiwan University, Taipei 106, Taiwan\\
}
\ead{sfujii@mst.nagaokaut.ac.jp}

\begin{abstract}
We study the elastic properties of thermotropic smectic liquid 
crystals with focal conic domains (FCDs). 
After the application of the controlled preshear at different 
temperatures, we independently measured the shear modulus $G'$ and 
the FCD size $L$. 
We find out that these quantities are related by the scaling relation 
$G'\approx \gamma_{\rm eff}/L$ where $\gamma_{\rm eff}$ is the 
effective surface tension of the FCDs.
The experimentally obtained value of $\gamma_{\rm eff}$ shows the 
same scaling as the effective surface tension of the layered systems
$\sqrt{KB}$ where $K$ and $B$ are the bending modulus and 
the layer compression modulus, respectively.
The similarity of this scaling relation to that of the surfactant 
onion phase suggests an universal rheological behavior of the 
layered systems with defects.
\end{abstract}

\submitto{\JPCM}

\section{Introduction}

The rheology of smectic liquid crystal is a long-standing
puzzle~\cite{Larson,OP,LWPWB}.
Although the theory predicts a liquid response when the smectic layers
are aligned parallel to the shear plane, some experiments showed that
the system behaves as a solid until certain critical shear stress is
exceeded~\cite{HK,COGCDGL,CNCO}.
This means that even the oriented smectic liquid crystal behaves as
a yield stress fluid at very low stress.
Such a peculiar behavior can be attributed to the presence of defects 
such as focal conic domains (FCDs) in the smectic layers, because they 
act to hinder layer sliding in the low stress limit.
Indeed it was seen that the FCD density strongly correlates with the 
viscoelastic properties~\cite{LWPWB,HK}.
Previously we showed that the yield stress remarkably decreases 
around the SmA-N transition temperature, and vanishes at the transition 
point~\cite{FIKL}.
These observations were attributed to the rapid increment of 
FCD size because the transition is induced by a smectic melting. 
It is known that the SmA-N transition of 8CB is very close to the second 
order one~\cite{BMNR,BCMR}. 
In general, softening of the elasticity in the smectic close to the 
transition point can be induced by defect unbinding~\cite{NT}.

Prior to our work, it was reported that defects of 
oily streaks stabilized by colloidal particles reinforce the shear 
modulus which increases with the defect density~\cite{BSKNRNR,RZLW}.
For 8CB liquid crystal, it was shown that defects that are artificially 
introduced by aerosil gel network result in the soft glassy nature of 
the smectic phase~\cite{BLCHL}.
Shear modulus of the smectic phase is strongly influenced by the line 
tension of dislocations because it acts against the 
Peach-Koehler force which drives the motion of dislocations.
The balance between these two forces determines the dislocation spacing 
which in turn affects the elasticity of the smectic phase. 
When the force exerted on the dislocations exceeds the line tension,
the system undergoes plastic deformation.
Meyer \textit{et al.} proposed a theory of smectic rheology associated 
with the motion of screw dislocations in the plastic deformation 
region~\cite{MABK2,MAK}.
Using the Orowan relation~\cite{Orowan}, they predicted that the shear 
stress $\sigma$ and the shear rate $\dot\gamma$ follows the scaling 
relation $\sigma \sim \dot\gamma^{3/5}$, and the dislocation spacing
$\xi$ decreases as $\xi\sim\dot\gamma^{-1/5}$.
Hence the smaller $\xi$ results in the higher shear modulus.

On the other hand, it is known that the smectic phase is occupied 
by FCDs under shear.
In thermotropic smectic liquid crystals without any additives,
the proliferation of dislocation loop is induced by both equilibrium 
thermal fluctuations and non-equilibrium shear flow~\cite{LCIKK,dGP}.
Horn and Kleman investigated the origin of the elasticity of the smectic 
phase by comparing the rheometry and the birefringence 
measurements~\cite{HK}.
Based on the dimensional analysis, they proposed a scaling relation 
for the yield stress $\sigma_{\rm y}$ such that 
$\sigma_{\rm y} \sim K/L^{2}$, where $L$ is the FCD size and $K$
is the bending modulus.
Although it is unknown how the dislocation unbinding affects the FCD 
size, the strong correlation between temperature and flow should be 
reflected on the shear modulus through the change in the FCD size.

The aim of this study is to investigate the origin of the 
elasticity in the smectic phase of a thermotropic liquid crystal (8CB)
by comparing the shear modulus and the FCD size obtained from 
independent measurements.
We discuss the preshear and the temperature dependences of the shear 
modulus, and relate it to the corresponding FCD size. 
The measured shear modulus is attributed to the energy cost 
for deformation of FCDs under shear.
Thus their surface tension which acts against the deformation 
plays an important role in determining the shear modulus. 
The effective surface tension of FCD is given by 
$\gamma_{\rm eff}=\sqrt{KB}$, where $K$ and $B$ are the bending modulus 
and the compression modulus, respectively.
Notice that $\gamma_{\rm eff}$ has the dimension of energy per area.
We demonstrate that the shear modulus is inversely proportional to the 
FCD size with the coefficient determined by the surface tension of FCDs.
A notable analogy with the elasticity of the surfactant onion 
phase will be also discussed, whereby suggesting an universal rheological
property of the layered structures with defects.

\begin{figure}
\centering
\includegraphics[scale=0.6]{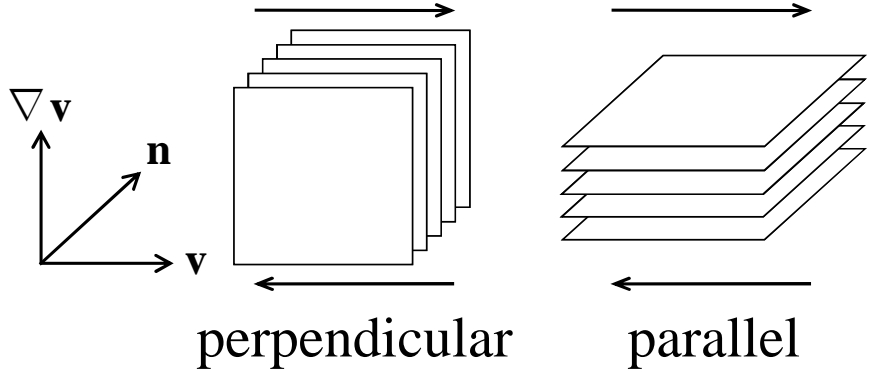}
\caption{Schematic diagram of a smectic phase with the perpendicular 
(left) and the parallel (right) orientations under shear flow.
$\nabla {\bf v}$, ${\bf v}$, and ${\bf n}$ correspond to the flow gradient, 
flow, and vorticity directions, respectively. 
The SmA$_{\rm I}$ region includes both the perpendicular and the parallel 
orientations, while only the perpendicular orientation appears in the 
SmA$_{\rm II}$ region~\cite{FIKL}.
}
\label{layer}
\end{figure}

In our previous study on the same system, we showed that the two 
different types of layer orientation exist in the dynamical phase 
diagram, i.e., SmA$_{\rm I}$ and SmA$_{\rm II}$~\cite{FIKL}.
In the SmA$_{\rm I}$ region, both the parallel and the perpendicular 
orientations of the smectic layers shown in figure~\ref{layer}
coexist as originally suggested by Safinya \textit{et al.}~\cite{SSP}.
On the other hand, the SmA$_{\rm II}$ phase is dominated by the 
perpendicular orientation in which the layer normal points along the
vorticity direction~\cite{SSP,PAR}.
The SmA$_{\rm I}$ phase transforms into the SmA$_{\rm II}$ phase upon
increasing either shear stress or temperature.
In this article, we are mostly concerned with the SmA$_{\rm I}$ regime,
because the rheological property of the SmA$_{\rm II}$ phase is simply 
Newtonian due to the perpendicular orientation of the smectic layers.

\section{Experimental}

We used the thermotropic liquid crystal 4-n-octyl-4'-cyanobiphenyl
(8CB) in the smectic phase. 
The equilibrium phase sequence and some physical properties of 8CB 
have been studied in the literatures~\cite{SSP,PAR}. 
8CB was obtained from SYNTHON Chemicals GmbH \& Co.\ Germany and was 
used without further purification. 
Rheological measurements were performed using the
stress-controlled rheometer, Anton Paar MCR-300, equipped with a
truncated cone-plate geometry whose diameter is 50 mm and the cone angle 
is 1 degree. 
In contrast to reference \cite{HK}, no surface treatment has been 
done in our experiment.
The absence of the homeotropic alignment of the smectic layers would 
easily induce the nucleation of focal parabolae under shear flow.
However, we have checked that the reproducible results can be 
obtained by applying the preshear even without any surface treatment. 
As we will show in next section, our preshearing causes a similar effect 
to the anchoring treatment.
Temperature was controlled within 0.02 K by a Peltier device attached 
to the rheometer. 
We determined the SmA-N transition temperature $T_{\rm SN}$ by the 
dynamic viscoelastic measurement.
From the condition that the storage modulus vanishes, we identified it
as $T_{\rm SN}=33.4$ $^\circ$C.
This is in good agreement with the previously reported value~\cite{SSP}. 
The samples were presheared before the dynamic measurement at 
different shear stresses such that the FCD density was well controlled.
Each shear stress was applied for 600 s which is long enough to reach
the steady state.

Microscope observation under shear flow was performed by using the 
Linkam CSS 450 shear cell which has a plate-plate shear geometry 
attached onto the Olympus microscope, BX-50, between the crossed 
polarizers. 
The gap size between the two plates was fixed to 150 $\mu$m. 
The microscope pictures were taken after applying the shear flow
for 600 s with different fixed shear rates.

\section{Results}

\subsection{Scaling of shear modulus}

\begin{figure}
\centering
\includegraphics[scale=0.6]{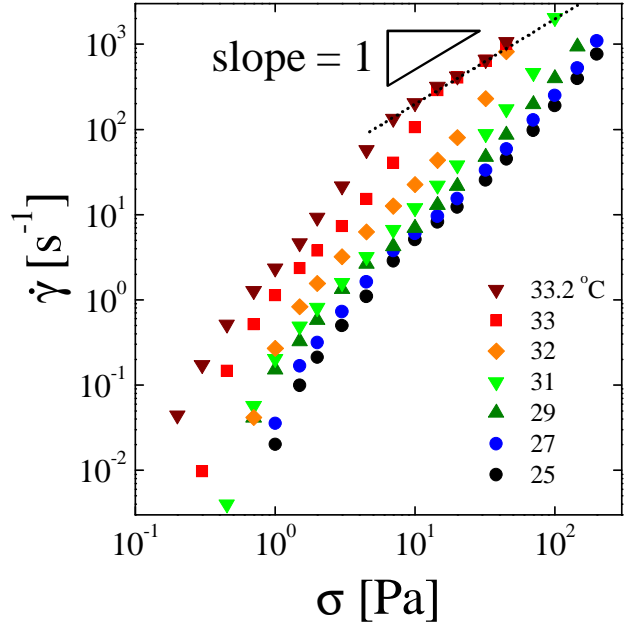}
\caption{
Steady state shear rate $\dot\gamma$ as a function of
the applied shear stress $\sigma$ at different temperatures.
The slope of the dotted line is unity showing the Newtonian 
behavior $\dot\gamma \sim \sigma$.
}
\label{flow}
\end{figure}

For each applied shear stress, the dynamic modulus was measured after 
the steady state shear rate was reached. 
The measured steady state shear rate $\dot\gamma$ as a function of the 
applied shear stress $\sigma$ is shown in figure \ref{flow}.
These data quantitatively reproduce our previous rheological 
measurement~\cite{FIKL}.
Most of the flow curves exhibit the shear thinning behavior.
At high temperatures such as $33.2$ $^\circ$C, a dynamical transition 
from the shear thinning behavior to the Newtonian behavior is observed 
for large applied shear stresses.
We note that the shear thinning and the Newtonian regimes correspond 
to the SmA$_{\rm I}$ and the SmA$_{\rm II}$ phases, respectively, as 
mentioned before. 
The data for $25$ $^\circ$C almost coincides with that obtained 
by Horn and Kleman who performed homeotropic anchoring 
treatment~\cite{HK}.
This implies that the our preshearing treatment causes an analogous
effect to the homeotropic anchoring treatment.
Hereafter, we shall mainly discuss the shear modulus of the 
SmA$_{\rm I}$ phase.

\begin{figure}
\centering
\includegraphics[scale=1]{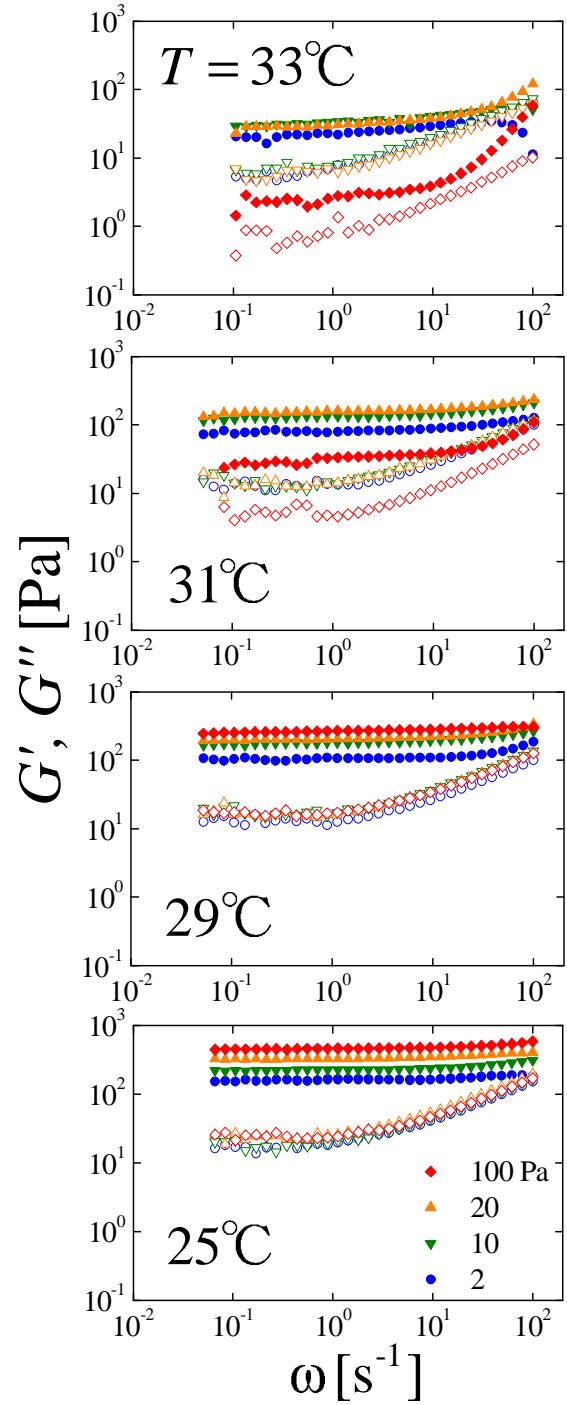}
\caption{
Dynamic storage modulus $G'$ and loss modulus $G''$ as a function
of frequency $\omega$ at different temperatures.
Filled and open symbols correspond to $G'$ and $G''$, respectively.
Different symbols shown in the bottom panel represent the applied 
preshear stresses for which the steady states are obtained. 
}
\label{modulus1}
\end{figure}

Figure~\ref{modulus1} shows the frequency dependence of the dynamic 
storage modulus $G'$ and the loss modulus $G''$ in the smectic phase
at different temperatures.
As reported before~\cite{COGCDGL}, we see that 8CB in the smectic phase 
generally shows a solid like behavior since $G'>G''$ for most of the
frequencies measured.
The extremely slow viscoelastic relaxation at low frequency regime,
indicative of the plastic behavior, can be attributed to the presence 
of defect structures.
Indeed, Larson \textit{et al.} showed that the elimination of defects 
using large amplitude oscillatory shear diminishes both $G'$ and 
$G''$~\cite{LWPWB}.

At each temperature, the plateau value of $G'$ becomes larger as 
the preshear stress $\sigma$ increases, while $G''$ does not show 
any remarkable dependence on it. 
However, a further increase of $\sigma$ caused a reduction of $G'$ 
and $G''$ as seen at $T=31$ and $33$ $^{\circ}$C in 
figure \ref{modulus1}.
The preshear dependence of $G'$ can be attributed to the variation 
in the defect density which is regulated by the 
preshearing process~\cite{LWPWB,HK}.
We also find that the plateau value of $G'$ decreases with increasing 
the temperature.
Especially, $G'$ becomes remarkably smaller close to the Sm-N transition
temperature $T_{\rm SN}$. 
Such a fall-off of $G'$ by increasing the temperature is analogous 
to the decrease of the yield stress of the same smectic phase 
as reported in reference \cite{FIKL}.

\begin{figure}
\centering
\includegraphics[scale=0.6]{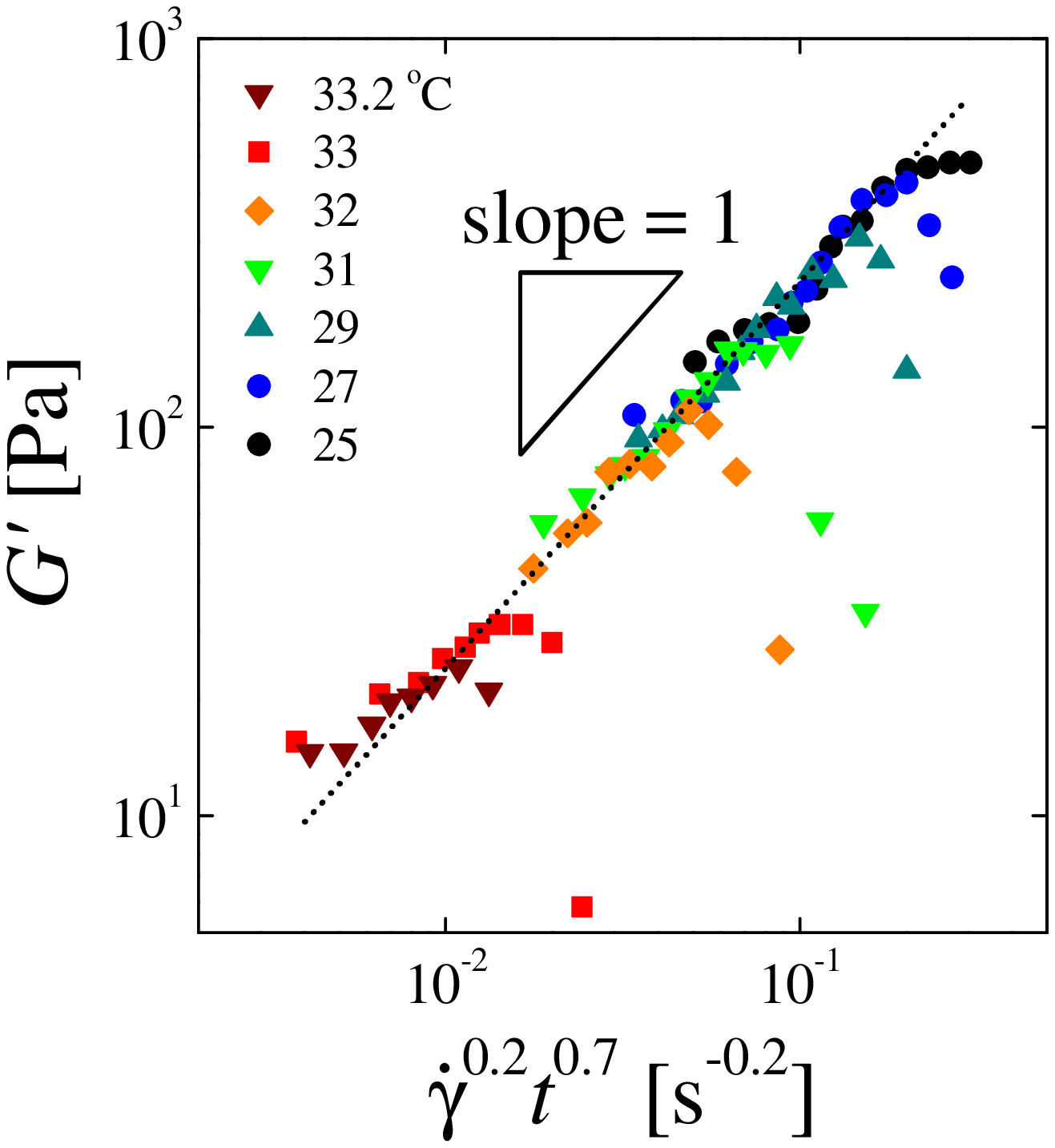}
\caption{
Plot of the plateau shear modulus $G'$ as a function of the shear 
rate $\dot\gamma$ and the reduced temperature $t$. 
Different symbols corresponds to different temperatures.
The scaling variable is chosen to be $\dot\gamma^{0.2}t^{0.7}$ 
so that most of the data points fall onto a straight dotted line whose 
slope is unity.
}
\label{scale2}
\end{figure}

Figure~\ref{scale2} presents the dependence of the plateau value of 
the storage modulus $G'$ at $\omega=0.1$ s$^{-1}$ on the shear rate 
$\dot\gamma$ and the reduced temperature $t=(T_{\rm SN}-T)/T_{\rm SN}$.
(In the analysis, we used the absolute temperature so that 
$T_{\rm SN}=306.5$ K.)
Adopting the result of figure \ref{flow}, we use here the measured 
steady state shear rate $\dot\gamma$ for each applied preshear stress 
$\sigma$.
To obtain the scaling plot of figure \ref{scale2}, we first determined 
the power law behavior $G'\sim \dot\gamma^{0.2}$ for each fixed 
temperature.    
Then we extracted the power law behavior of the scaled $G'$ as 
a function of $t$ with an exponent 0.7 so that all the data points 
fall onto a single straight line whose slope is unity.
After these procedures, we find the final scaling behavior 
$G' \sim \dot\gamma^{0.2} t^{0.7}$.
Notice that the slope of the dotted line in figure \ref{scale2} 
is unity. 
The data collapse is satisfactory except for those at high preshear 
stresses where $G'$ decreases.
The data which do not follow the scaling behavior indicate that 
they are close to the border between the SmA$_{\rm I}$ and 
SmA$_{\rm II}$ phases. 
This result indicates that the elasticity of the smectic phase 
is more enhanced as the shear rate (or the preshear stress) is 
increased and/or the temperature is lowered away from the 
transition temperature.
The fact that $G'$ can be scaled by the combined variable 
$\dot\gamma^{0.2} t^{0.7}$ means that both the shear rate and 
the proximity to the SN transition temperature have effectively 
the same influence on the elasticity of the smectic phase.
For the onion phase in the surfactant lamellar phase, 
the dependence of the yield stress on the shear rate is also 
given by the power law behavior with the same exponent 0.2 
in spite of the different structures~\cite{FW}.

\subsection{Scaling of FCD size}

\begin{figure}
\centering
\includegraphics[scale=1]{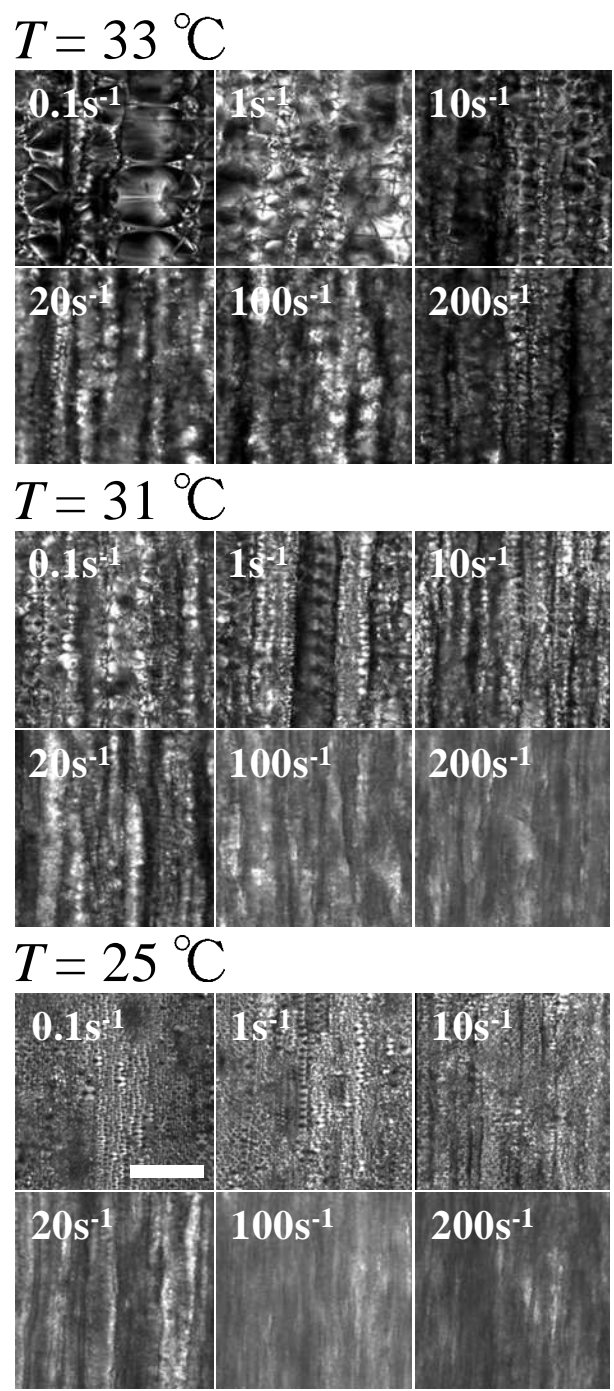}
\caption{
Polarized microscope images of the smectic phase under shear 
flow at different temperatures ($T =25, 31$, and $33$ $^\circ$C)
and shear rates ($\dot \gamma=0.1, 1, 10, 20, 100$, and 
$200$ s$^{-1}$). 
The flow is applied along the longitudinal direction.
The scale bar corresponds to 100 $\mu$m.
}
\label{microscope}
\end{figure}

Here we discuss the real space observation of the smectic textures 
under the optical microscope.
Figure~\ref{microscope} presents polarized microscope images obtained
after applying preshear at different temperatures for various 
fixed shear rates $\dot\gamma$.
These microscope images were obtained by observing the sample 
along the flow gradient direction $\nabla {\bf v}$.
In these images, the flow direction ${\bf v}$ and the vorticity 
direction ${\bf n}$ correspond to the longitudinal and the horizontal 
ones, respectively.
These pictures show that the FCDs occupy most of the observed area. 
For each temperature, the FCDs are made smaller with increasing $\dot\gamma$.
If one further raises $\dot\gamma$, the FCDs become too small 
to be observed under the optical microscope.
For example, FCDs are invisible in the image of $T=25$ $^\circ$C 
and $\dot\gamma=200$ s$^{-1}$.
We consider that the shrinkage of the FCDs is related to the
drop-off of $G'$ at high preshear stresses in figure~\ref{modulus1}.
Hence the annihilation of FCDs in the microscope images
and the deviation of $G'$ from the dotted line in 
figure \ref{scale2} is a signal for the onset of the 
parallel-to-perpendicular orientation transition.
As for the temperature dependence, the FCD size remarkably grows up 
with increasing temperature towards the transition temperature 
$T_{\rm SN}=33.4$ $^\circ$C.
Such a trend can be clearly seen by comparing the smectic textures
at $\dot\gamma=0.1$ s$^{-1}$ for different temperatures.

\begin{figure}
\centering
\includegraphics[scale=0.6]{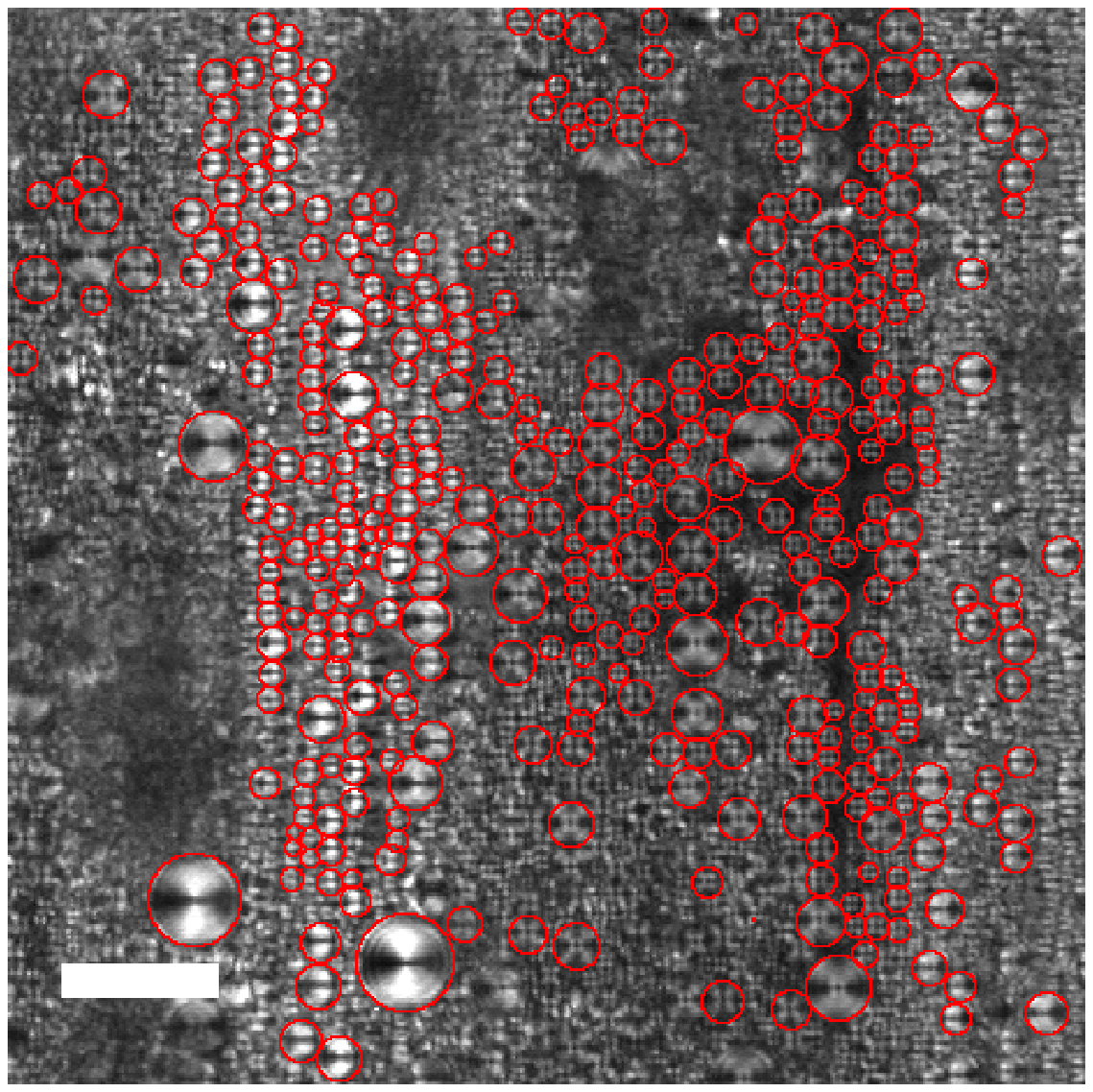}
\caption{Typical microscope image used to estimate the FCD size at 
$T=25$ $^\circ$C and $\dot{\gamma}=0.1$ s$^{-1}$.
The averaged diameter $L$ of FCDs was estimated by tracing each FCD with 
a red circle as shown in the picture.
The scale bar corresponds to 50 $\mu$m.
}
\label{micro}
\end{figure}

A typical example to extract the FCD size is presented in 
figure~\ref{micro} which is obtained at $T=25$ $^\circ$C when the 
shear rate is $\dot{\gamma}=0.1$ s$^{-1}$. 
As shown in figure~\ref{micro}, the outlines of the FCDs were traced 
with circles on the microscope images. 
Applying the same procedure to several microscope images obtained under
the same condition, we extracted the averaged diameter $L$ for the 
given temperature and the shear rate.
Although the FCD size obeys the distribution of the random Apollonian 
packing~\cite{RROC}, we have extracted the averaged diameter $L$ by 
ignoring rather small FCDs. 
However, our result is essentially independent of the way we evaluate 
the average FCD size besides some numerical factors.

\begin{figure}
\centering
\includegraphics[scale=0.6]{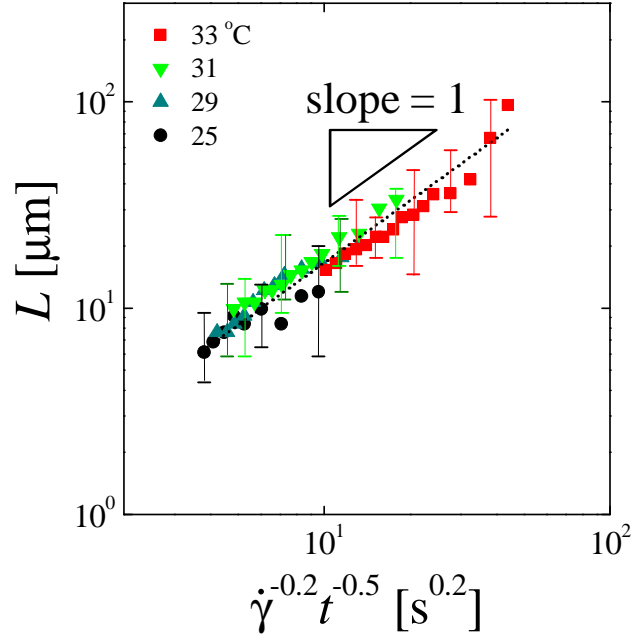}
\caption{
Plot of the FCD size $L$ as a function of the shear 
rate $\dot\gamma$ and the reduced temperature $t$. 
Different symbols corresponds to different temperatures.
The scaling variable is chosen to be $\dot\gamma^{-0.2}t^{-0.5}$ 
so that all the data points fall onto a straight dotted line whose 
slope is unity.
}
\label{scaleL}
\end{figure}

In figure \ref{scaleL}, we summarize the shear rate and the temperature 
dependence of the average FCD size. 
Here the upper and lower limits of the error bars correspond 
to the maximum and minimum size of the FCD, respectively.
Our analysis revealed that $L$ obeys the power law behaviors 
on the shear rate $\dot\gamma$ as $L \sim \dot\gamma^{-0.2}$,
and on the reduced temperature $t$ as $L \sim t^{-0.5}$.
Hence the average FCD size can be well scaled by the combined
variable $\dot\gamma^{-0.2}t^{-0.5}$ as verified in 
figure \ref{scaleL}. 
Here the data collapse onto a single line of slope unity is clearly 
demonstrated so that we can deduce the scaling relation  
$L \sim \dot\gamma^{-0.2}t^{-0.5}$. 
To the best of our knowledge, there is no model which 
describes the shear rate dependence of the FCD size. 
However, it is worthwhile to point out that a similar scaling 
relation $\xi\sim\dot\gamma^{-1/5}$ was predicted between the 
defect spacing $\xi$ and the shear rate $\dot\gamma$ as mentioned in 
Introduction~\cite{MABK2,MAK}.
Generally, FCDs can form strings connected by edge 
dislocations~\cite{BLK}, while a dislocation loop consists of 
pairs of screw and edge dislocations.
Although the topological relation between FCDs and dislocation 
loops is unknown, the same exponent 0.2 may arise provided that 
the scale of edge dislocations regulate the FCD size.

Regarding the temperature dependence, the following discussions
are in order. 
According to the defect model of the SmA-N transition proposed by
Helfrich~\cite{Helfrich}, smectic layers are destroyed by  
proliferation of dislocation loops as the transition point is 
approached from below.
He predicted that the average size of the dislocation loops  
diverges at the SmA-N transition point as    
$\xi \sim t^{-0.5}$.
Such a melting behavior of the smectic order was experimentally 
confirmed in the vicinity of the smectic-cholesteric 
transition for a chiral system~\cite{MNGMNR}.
If we assume that there exists only one characteristic length 
scale close to the critical temperature~\cite{dGP}, $\xi$ should 
be proportional to the FCD size $L$, and hence the obtained relation 
$L\sim t^{-0.5}$ is reasonable.

\subsection{Origin of elasticity}

By conducting independent measurements, 
we have experimentally obtained the scaling relation 
for the shear modulus as $G' \sim \dot\gamma^{0.2}t^{0.7}$ 
and that for the average FCD size as  
$L \sim \dot\gamma^{-0.2}t^{-0.5}$ .
These two relations strongly lead us to suggest that $G'$ is inversely 
proportional to $L$, although this seems to be not completely 
correct because of the slight difference in the temperature 
exponents (0.7 vs.\ 0.5).
However, this inconsistency can be resolved if the proportionality 
coefficient would also depend on the temperature.
We show below that this is indeed the case.

If we assume that the relation $G'\sim 1/L$ holds, the 
proportionality coefficient in the right hand side should have the 
dimension of surface tension, i.e., energy per area.     
De Gennes pointed out that the layered system such as smectic 
phase or lamellar phase indeed exhibits the surface 
tension~\cite{dGP}.      
According to the references \cite{dGP,vdL,LHL}, the effective surface 
tension of the layered system is given by 
$\gamma_{\rm eff}=\sqrt{KB}$ where $K$ and $B$ are the bending 
modulus and the compression modulus of the smectic phase, respectively.
We then propose the following relation 
\begin{equation}
G' = C \frac{\sqrt{KB}}{L},
\label{eq1}
\end{equation}
where $C$ is a dimensionless numerical coefficient.
From the previous experimental papers, it is known that $K$ for 8CB 
is almost constant $K=(5.2 \pm 0.3) \times 10^{-12}$ N~\cite{ZPOG},
while $B$ does depend on the temperature close to $T_{\rm SN}$
obeying the power law relation   
$B=(7.5 \times 10^{7}) \cdot t^{0.4 \pm 0.03}$ Pa~\cite{BMNR}.
With this relation, $B$ vanishes at the transition point ($t=0$)
because the SmA-N transition is associated with the disruption 
of the layered structures.

In order to check the validity of (\ref{eq1}), we compare
the values of $G'L$ obtained from our experiment and 
$\gamma_{\rm eff}=\sqrt{KB}$ estimated from the literatures.        
By taking into account the prefactors in our previous scaling 
relations, we obtain $G'L = (4.56 \times 10^{-3}) \cdot t^{0.2}$ N/m.
On the other hand, the above values from the literatures yield
$\sqrt{KB} = (1.97 \times 10^{-2}) \cdot t^{0.2}$ N/m.
Hence (\ref{eq1}) is totally valid in terms of the scaling 
with respect to both $\dot\gamma$ and $t$.
Moreover, the dimensionless numerical factor can be determined 
to be $C=0.456/1.97 \approx 0.23$.
The consistency of (\ref{eq1}) demonstrates that the physical 
origin of the elasticity in the smectic phase is due to the 
effective surface tension $\gamma_{\rm eff}$ of the FCDs.
This is the main claim of the present paper.

Princen \textit{et al.} experimentally showed that the physical origin 
of the shear modulus and the yield stress in emulsions can be commonly 
attributed to the surface tension, although the volume fraction 
dependences of the shear modulus and the yield stress are different 
each other~\cite{PK86,PK89}. 
The fact that the above prefactor $C$ is smaller compared to that 
of monodispersed emulsions can be attributed to the polydispersity 
of the FCD size~\cite{PK86,PK89}.

\section{Discussion}

We first discuss the generality of (\ref{eq1}).
In fact, essentially the same relation was proposed for the 
shear-induced onion phase in surfactant 
solutions~\cite{PRVLC}.
They experimentally found that the shear modulus of the onion
phase follows the relation
$G'= C_0 + C_1 \sqrt{KB}/L$. 
Here the first constant term $C_0$ reflects the residual strain 
due to the disordered arrangement of the onion structure.
The second term is the stored energy per unit volume, and the 
quantity $\sqrt{KB}$ corresponds to the effective surface 
tension of a single onion under small deformation~\cite{vdL,LHL}.
Hence they concluded that the elasticity of the onion phase 
originates in the energetic penalty associated with the onion 
deformation.
The later experimental work reported that $C_0$ tends to vanish 
for the disordered onion systems, while the prefactor $C_1$ 
varies from 0.4 to 1.2 for different surfactants~\cite{LNR}.
These results are in accord with our result for the smectic phase 
with FCDs.

It is well known that the structure of FCD is classified into 
two types; one is the toroidally deformed FCD (FCD-I) commonly 
observed in the thermotropic smectics, and the other is the 
onion type (FCD-II) seen in the lyotropic liquid crystals~\cite{OP}.
In spite of the topological difference between the two types,
our result suggests that the energy cost for the deformation is 
determined by $\sqrt{KB}$ both for FCD-I and FCD-II. 
Both in the smectic phase and the surfactant onion phase, the 
main source of the elasticity seems to be the effective surface 
tension.
Therefore the rheological response of these systems is generally 
described by the scaling relation of (\ref{eq1}) which is universal.

As mentioned before, Horn and Kleman suggested a scaling relation 
$\sigma_{\rm y} \sim K/L^{2}$  
for the yield stress of the smectic phase with FCDs~\cite{HK}. 
This relation was also used in our previous paper in order to 
estimate the FCD size close to the transition 
temperature~\cite{FIKL}. 
(A similar relation also holds even for the elasticity due 
to the line defects when $K$ and $L$ are replaced by the line tension 
and the defect spacing, respectively, because the elasticity of the 
defect network is proportional to the yield 
stress~\cite{BSKNRNR,RZLW}.)
We emphasize that our result (\ref{eq1}) differs from it.
We now speculate that the elasticity of the layered structures 
with FCDs should be properly accounted for by both $K$ and $B$. 
Especially, the layer compression modulus $B$ reflects the
overall interactions between the adjacent layers.

It should be noted that the maximum size of the FCD at 
$\dot\gamma=0.1$ s$^{-1}$ and $T=33$ $^\circ$C is about $100$ $\mu$m 
which is smaller than the gap size $150$ $\mu$m of the plate-plate cell.
Hence we consider that the effect of the gap size on the FCD size is 
almost negligible.
For the viscoelastic measurement in which the gap size changes from 
$50$ to $436$ $\mu$m, the FCDs at higher temperatures and lower shear 
rates would be compressed close to the center of the cone.
However, the volume of such a compressed FCD region is negligibly small 
compared with the total volume of the sample. 
We thus believe that the gap size effect on the FCD size will not produce 
any significant influence on the shear modulus.

In our analysis, we have used the experimentally observed 
scaling $B \sim t^{0.4}$~\cite{BMNR}.
However, the real critical behavior of $B$ requires a more careful 
discussion because it was reported later that $B$ does not 
necessarily vanish at the transition point~\cite{BCMR}. 
This result supports well the Nelson-Toner model of the 
SmA-N transition with full anisotropy~\cite{NT}. 
In our experiment, the temperature is not very close to the 
critical temperature so that the above relation is practically
useful. 
On the other hand, a caution is required when we use this 
relation beyond the critical region in the low temperature.

\section{Conclusion}

In summary, we have studied the viscoelastic properties of 
thermotropic smectic liquid crystals with FCDs.
By measuring the shear modulus $G'$ and the FCD size $L$ independently,
we have found the scaling relations 
$G' \sim \dot\gamma^{0.2} t^{0.7}$ and 
$L \sim \dot\gamma^{-0.2}t^{-0.5}$.
The product $G'L \sim t^{0.2}$ scales in the same manner as the 
effective surface tension $\gamma_{\rm eff}=\sqrt{KB} \sim t^{0.2}$.
Our experimental finding suggests that the shear modulus is inversely 
proportional to the FCD size and the proportionality coefficient is 
set by the effective surface tension $\gamma_{\rm eff}$.
Hence we conclude that the physical origin of the elasticity in the 
smectic phase is attributed to the effective surface tension of FCD.
The similarity of the rheological response between the smectic phase
and the surfactant onion phase points toward the universal behavior 
of the layered systems under non-equilibrium conditions.


\ack
We acknowledge useful discussions with R.\ H.\ Colby.
We thank T.\ Takahashi and M.\ Imai for allowing us to use the 
rheometer MCR-300 and the Linkam shear cell CSS-450, respectively.
SF and SK acknowledge support by KAKENHI (Grant-in-Aid for Scientific
Research) on Priority Areas ``Soft Matter Physics'' and Grant
No.\ 21540420 from the Ministry of Education, Culture, Sports,
Science and Technology of Japan.

\section*{References}


\end{document}